\documentclass[fleqn,usenatbib]{mnras}

\usepackage{newtxtext,newtxmath}

\usepackage[T1]{fontenc}
\usepackage{ae,aecompl}

\usepackage{graphicx}	
\usepackage{amsmath}	
\usepackage{amssymb}	
\usepackage{float}
\usepackage{subfigure}
\usepackage{gensymb}

\newcommand{\HI}{{\sc H\,i }}

\title[Mapping the Dark Matter Halo of NGC 2974]{Mapping the dark matter halo of early-type galaxy NGC 2974 through orbit-based models with combined stellar and cold gas kinematics}

\author[Yang et al.]{Meng Yang$^{1}$\thanks{E-mail: my38@st-andrews.ac.uk},
Ling Zhu$^{2}$,
Anne-Marie Weijmans$^{1}$, 
Glenn van de Ven$^{3,4}$, \newauthor
Nicholas Boardman$^{5}$,
Raffaella Morganti$^{6,7}$,
Tom Oosterloo$^{6,7}$
\\
$^{1}$School of Physics and Astronomy, University of St Andrews, North Haugh, St Andrews, KY16 9SS, UK\\
$^{2}$Shanghai Astronomical Observatory, Chinese Academy of Sciences, 80 Nandan Road, Shanghai 200030, China\\
$^{3}$Department of Astrophysics, University of Vienna, Türkenschanzstrasse 17, 1180 Vienna, Austria\\
$^{4}$European Southern Observatory, Karl-Schwarzschild-Str 2, D-85748 Garching bei Munchen, Germany\\
$^{5}$Department of Physics and Astronomy, University of Utah, Salt Lake City, UT 84112, United States\\
$^{6}$Netherlands Institute for Radio Astronomy (ASTRON), Postbus 2, NL-7990 AA Dwingeloo, the Netherlands\\
$^{7}$Kapteyn Astronomical Institute, University of Groningen, PO Box 800, NL-9700 AV Groningen, the Netherlands\\
}

\date{Accepted XXX. Received YYY; in original form ZZZ}

\pubyear{2019}

\begin{document}
\label{firstpage}
\pagerange{\pageref{firstpage}--\pageref{lastpage}}
\maketitle

\begin{abstract}

We present an orbit-based method of combining stellar and cold gas kinematics to constrain the dark matter profile of early-type galaxies. We apply this method to early-type galaxy NGC 2974, using Pan-STARRS imaging and SAURON stellar kinematics to model the stellar orbits, and introducing  \HI kinematics from VLA observation as a tracer of the gravitational potential. The introduction of the cold gas kinematics shows a significant effect on the confidence limits of especially the dark halo properties: we exclude more than $95\%$ of models within the $1-\sigma$ confidence level of Schwarzschild modelling with only stellar kinematics, and reduce the relative uncertainty of the dark matter fraction significantly to $10\%$ within $5 R_\mathrm{e}$. Adopting a generalized-NFW dark matter profile, we measure a shallow cuspy inner slope of $0.6^{+0.2}_{-0.3}$ when including the cold gas kinematics in our model. We cannot constrain the inner slope with the stellar kinematics alone.
\end{abstract}

\begin{keywords}
galaxies:structure -- galaxies:kinematics and dynamics -- galaxies: haloes -- dark matter
\end{keywords}



\section{Introduction}
Dark matter haloes are not only crucial for investigating the nature of dark matter and testing cosmological models, but for galaxy formation and evolution as well. Galaxies form in dark matter haloes, and the accumulation of baryons reshapes the dark matter haloes. Thus, studying the structure of dark matter haloes is a way to understand the co-evolution processes of dark matter and baryons in galaxies. 

Several questions related to the dark matter structure are still under debate, for example, whether dark matter haloes are cusped or cored. N-body simulations show that the standard $\Lambda$CDM model \citep[e.g.][]{blumenthal1984formation,davis1985evolution} predicts dark matter haloes to have steep inner slopes called cusps, which are well described by a Navarro-Frenk-White profile \citep[NFW;][]{navarro1996structure}, while observations of dwarf spheroidal galaxies prefer shallow core-like inner slopes which could indicate warm dark matter particles \citep{moore1994evidence,moore1999cold,battaglia2008kinematic,walker2011method}. This question is still undetermined \citep{zhu2016discrete2}, and it requires further examination. 

The dark matter is expected to dominate the gravitational potential in the outer regions in galaxies. Spatially resolved stellar kinematics obtained with Integral Field Spectroscopy (IFS) observations, such as the Atlas$\rm ^{3D}$ Survey \citep{cappellari2011atlas3d}, CALIFA \citep{sanchez2012califa}, SAMI \citep{croom2012sydney} and MaNGA \citep{bundy2014overview}, have been widely used to trace mass distribution of galaxies \citep[e.g][]{cappellari2013atlas3d, li2017sdss, taranu2017self, zhu2018morphology}.  However, these stellar kinematics only have limited coverage of $1-2$ effective radius ($R_\mathrm{e}$), as the outer regions of galaxies are faint, which makes them difficult to observe. 
Other tracers extending out to over $5 R_\mathrm{e}$ in galaxies, such as planetary nebulae (PNe), globular clusters (GCs) and cold gas, are of great importance. 
Since \citet{hui1993planetary} first reported the possible existence of dark matter halo measured with PNe kinematics, PNe have been used to measure dark matter distribution in early-type galaxies \citep[e.g.][]{tremblay1995planetary,napolitano2007dark,napolitano2009planetary}. GCs are also good tracers for early-type galaxies \citep[e.g.][]{cote2001dynamics,cote2003dynamics,zhu2014next,alabi2017sluggs} because of their ubiquity and adequate luminosity for spectroscopic observation to a far distance \citep{norris2012globular}. 

Unlike discrete tracers such as PNe and GCs, cold gas is a continuous tracer following the intrinsic shape of the gravitational potential. There is a long history of ascertaining the dark matter content of late-type (spiral) galaxies with cold gas (typically neutral hydrogen \HI), by modelling rotation curves obtained from integrated line profiles and velocity fields \citep[e.g.][]{bosma198121,van1985distribution}. \HI discs are also present in early-type galaxies although typically with lower surface brightness as in late-type galaxies \citep[e.g.][]{morganti1997h,oosterloo2007extended,serra2012atlas3d}. NGC 2974 is one of the well-studied early-type galaxies with cold gas: it is a lenticular galaxy with an extended regular \HI ring \citep{kim1988hi,weijmans2008shape}. 

Multiple dynamical modelling techniques have been developed to reconstruct the gravitational potential of galaxies and detect their dark matter structure. Jeans models \citep{jeans1922motions} are applicable for integrable systems with a distribution function (DF) depending on phase-space coordinates only through integrals of motion \citep{binney2011galactic}, while the particle-based made-to-measure (M2M) algorithm \citep[e.g.][]{syer1996made,de2007nmagic} and the Schwarzschild's orbit-superposition technique \citep{schwarzschild1979numerical,van2008triaxial} regard the DF as a large ensemble of $\delta$-functions and sidestep the ignorance of integrals of motion.

Even though the dynamical modelling with extended stellar kinematics (to $ \sim 4 R_\mathrm{e}$) are able to constrain the dark matter profile in a number of cases \citep[e.g][]{forestell2010hobby,cappellari2015small,boardman2016low}, the combination of central stellar kinematics (within $\sim 1 R_\mathrm{e}$) and other extended tracers has expanded the methods to break the degeneracies in galaxies between dark and luminous matter. The combination of stellar kinematics and discrete tracers with the M2M method \citep[e.g.][]{de2008dark,das2011using,morganti2013elliptical} and the Jeans modelling \citep[e.g.][]{napolitano2011pn,napolitano2014sluggs,zhu2016discrete,bellstedt2018sluggs} have provided crucial measurements of the slope of the dark matter profile.
However, the similar combination still lacks application within Schwarzschild's modelling technique, which is usually unable to obtain much information of the dark matter without extended tracers, whether modelling individual galaxies including NGC 2974 \citep{krajnovic2005dynamical} or galaxy populations \citep[e.g.][]{zhu2018stellar}. Therefore, we introduce an combination of stellar kinematics modelled by the Schwarzschild technique, and cold gas kinematics as a tracer of the gravitational potential at large radii. We demonstrate this technique by applying it to NGC 2974, to obtain the dark matter properties of this galaxy. 

The organization of this paper is as follows: in Section 2, we introduce the data used in our dynamical models and in Section 3 we describe our dynamical modelling method in detail. In Section 4 we show our resulting models and make a comparison between those models with and without cold gas constraints. We discuss our results and their implications further in Section 5 and we summarize our work in Section 6.

\section{Data}
\label{sec:data}

To construct our dynamical models of NGC 2974, we make use of a variety of datasets, that we describe below. We list the basic properties of NGC 2974 in Table~\ref{tab:1}.

\begin{table}
\centering
\begin{tabular}{lr}
\hline
Parameter      & Value       \\
\hline
Hubble type      & S0$^{a}$       \\
Distance         & 20.89 Mpc \\
Distance scale   & 101.3 pc/arcsec \\
Position Angle & 41\degree \\
Effective radius ($R_\mathrm{e}$) & 24 arcsec \\
K-band magnitude ($M_\mathrm{K}$) & -23.62 mag \\
Effective stellar velocity dispersion ($\sigma_\mathrm{e}$) & 226 km/s \\
\hline
\end{tabular}
\caption{Basic Properties of NGC 2974. $^a$ NGC 2974 was firstly classified as an E4 galaxy, and then  \citet{cinzano1994observations} found it to be a lenticular (S0) galaxy. All other values were taken from \citet{weijmans2008shape}, except $M_\mathrm{K}$ \citep{cappellari2011atlas3d} and $\sigma_\mathrm{e}$ \citep{cappellari2013atlas3d}.
}
\label{tab:1}
\end{table}

\subsection{Surface Brightness}
To trace the stellar mass, we model the surface brightness of NGC 2974 based on $r$-band imaging taken from the Panoramic Survey Telescope and Rapid Response System (Pan-STARRS). The Pan-STARRS images are stacked from short exposure images, to reach a limiting magnitude of 23.2 in $r$-band. For further information on the Pan-STARRS, we refer the readers to \citet{chambers2016pan} and references therein. In NGC 2974, this correspond to a radius of 3.5 $R_\mathrm{e}$ (effective or half-light radii), which is beyond the extent of our stellar kinematic data sets. The Pan-STARRS $r$-band filter is comparable to the $r$-band filter of the Sloan Digital Sky Survey (SDSS) \citep{gunn20062,doi2010photometric}. 

\begin{figure}
	\centering
   	\subfigure[r-band image]{
	\label{image-1}
	\includegraphics[width=0.4425\columnwidth]{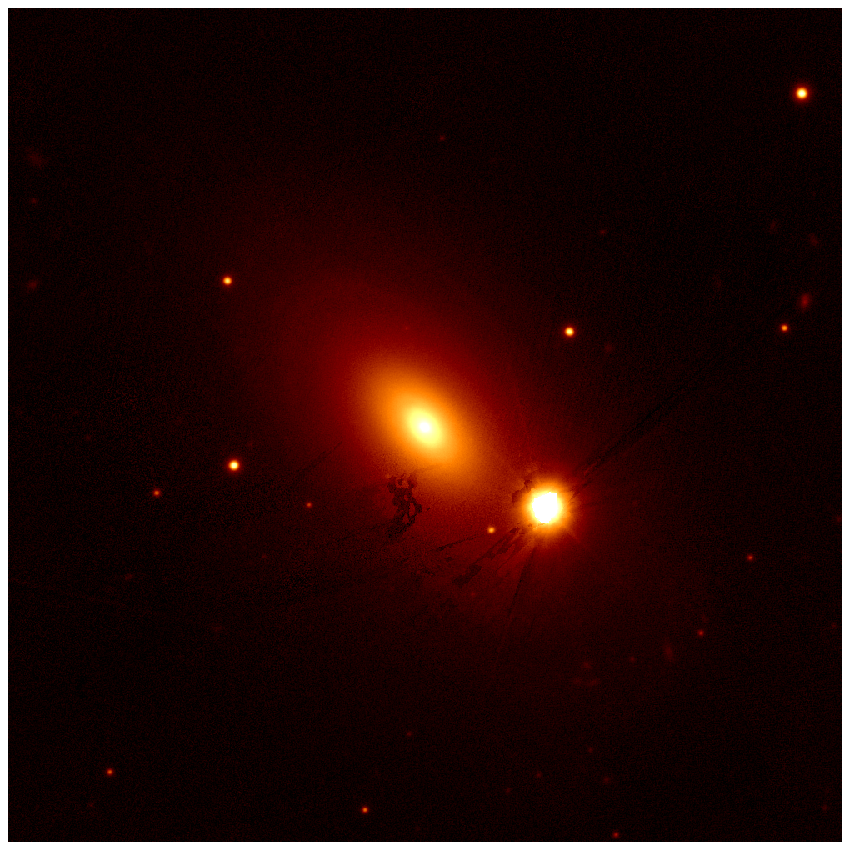}}
	\subfigure[contour map]{
	\label{image-2}
	\includegraphics[width=0.50\columnwidth]{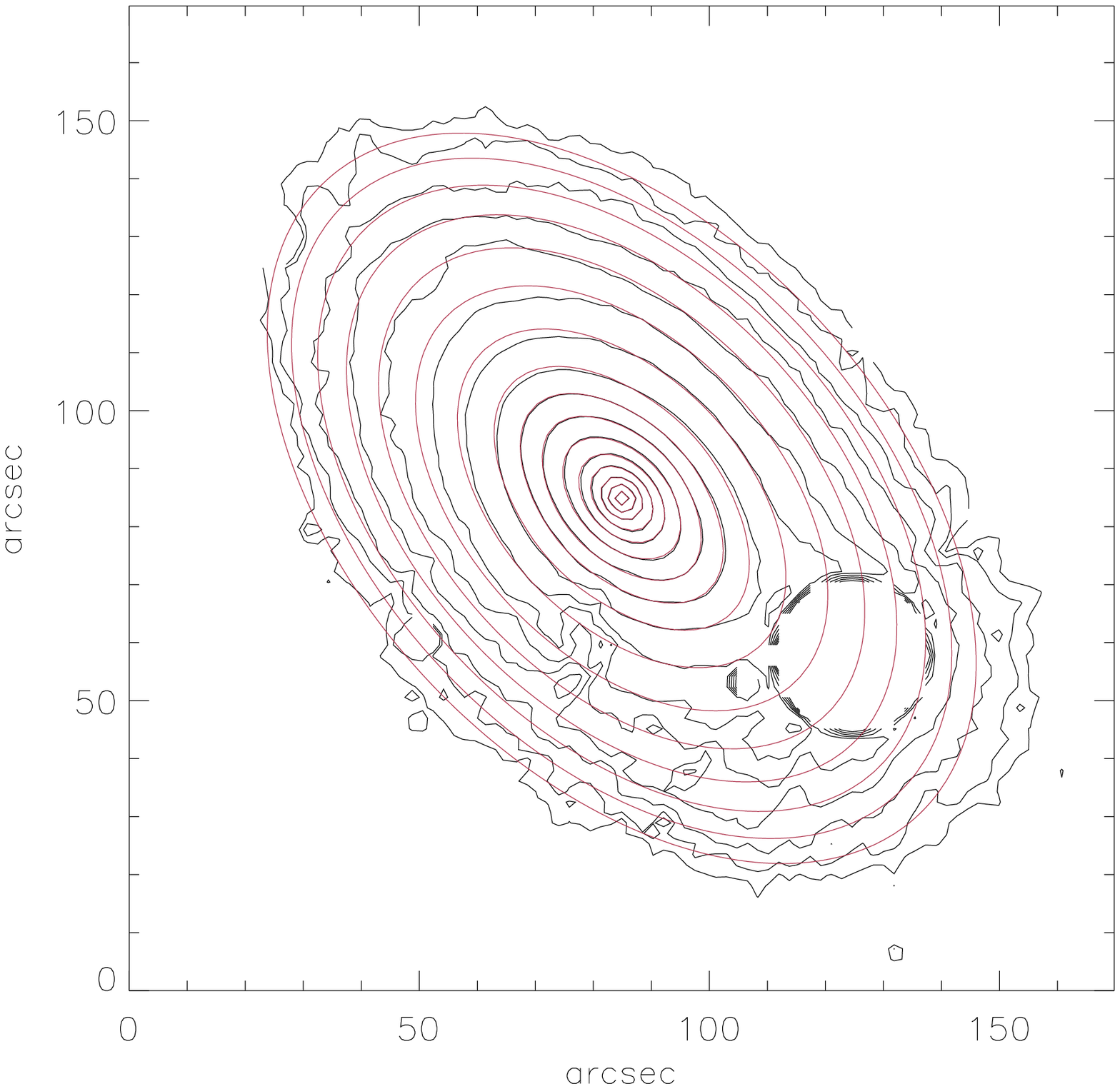}}
    \caption{(a) The $r$-band image of NGC 2974 from Pan-STARRS; (b) The surface brightness contours of NGC 2974 (black) and its Multi Gaussian Expansion (MGE) model (red). For a description of the MGE modelling method, see Section~\ref{subsec:MGE}. }
    \label{fig:image}
\end{figure}

In Figure~\ref{fig:image} we show the Pan-STARRS image of NGC 2974, as well as the resulting surface brightness model based on Multiple Gaussian Expansion (MGE) fitting, see Section~\ref{subsec:MGE}.

\subsection{Stellar Kinematics}
\begin{figure}
	\centering
	\includegraphics[width=\columnwidth]{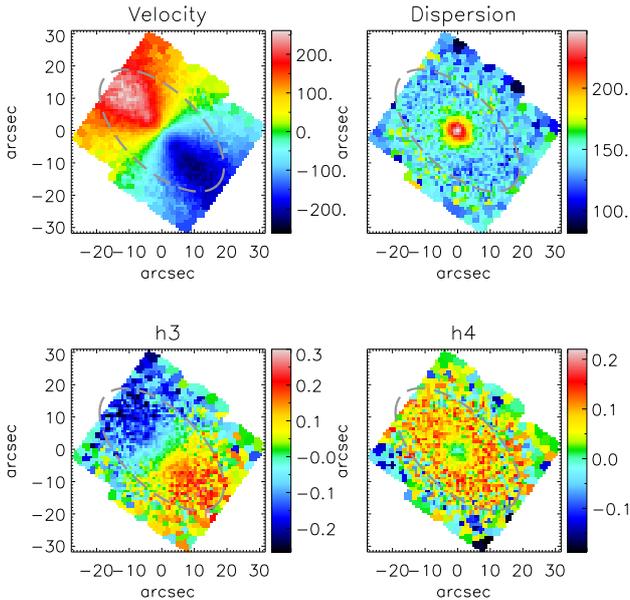}
    \caption{The stellar kinematic maps of NGC 2974 observed with SAURON, including velocity (km/s), velocity dispersion (km/s), the third and fourth orders of Gauss-Hermite moments. The maps are orientated so that north is up and east is to the left-hand side. The 1-$R_\mathrm{e}$ ellipse is plotted in dashed grey line.}
    \label{fig:stellar}
\end{figure}

NGC 2974 was observed with the SAURON integral-field unit on the William Herschel Telescope \citep{bacon2001sauron}, as part of the SAURON survey \citep{dezeeuw2002sauron}. The stellar kinematics (velocity, velocity dispersion and Gaussian-Hermite moments $h_3, h_4$ were first presented by  \citet{emsellem2004sauron}, while subsequently these observations were re-reduced as part of the Atlas$\rm ^{3D}$ Survey \citep{cappellari2011atlas3d}. In this work, we use the kinematic maps as published by Atlas$\rm ^{3D}$\footnote{www.purl.org/atlas3d}, see Figure~\ref{fig:stellar}. These kinematics were obtained using the penalised pixel fitting method pPXF\citep{cappellari2004parametric}, on spectra that were Voronoi binned\citep{cappellari2003adaptive} to a signal-to-noise of 40. More details on the extraction of the stellar kinematic are given in \citet{cappellari2011atlas3d}.

\subsection{Cold Gas Kinematics}
\begin{figure}
	\centering
	\includegraphics[width=0.8\columnwidth]{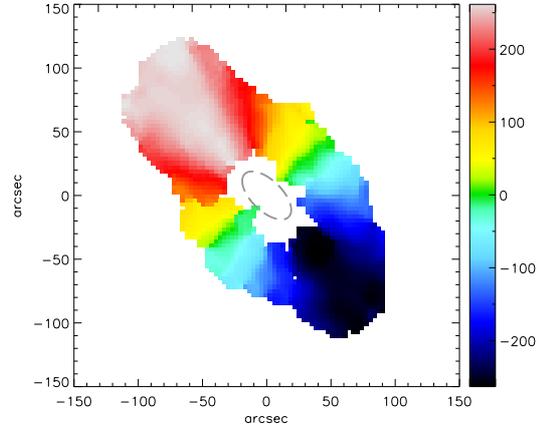}
    \caption{The \HI velocity map (km/s) of NGC 2974 observed with the VLA. The map is orientated so that north is up and east is to the left. The \HI ring is aligned with the stellar disc (see Figure~\ref{fig:stellar}). The 1-$R_\mathrm{e}$ ellipse is plotted in dashed grey line.}
    \label{fig:hi}
\end{figure}

We use \HI observations presented by \citet{weijmans2008shape}. These observations were obtained by the Very Large Array (VLA) in C-configuation, in September 2005. The data were reduced and calibrated using the MIRIAD software package \citep{sault1995retrospective}, resulting in a data cube with spectral resolution 20 km/s and a spatial beam of $19.9 \times 17.0$ arcsec$^2$. For more details on the \HI observations and data reduction we refer to \citet{weijmans2008shape}. We show the resulting \HI velocity map of NGC 2974 in Figure~\ref{fig:hi}: note that the stellar and cold gas discs are kinematically aligned. 

\section{Method}
\label{sec:method}
In this section, we describe our method of constraining the gravitational potential with the combination of two tracers: stars and cold gas. Stellar kinematics are modelled with an orbit-superposition Schwarzschild model, while cold gas kinematics are modelled as an ideal thin ring aligned with the stellar disc. First, we define a gravitational potential for our galaxy based on a choice of parameters (e.g., stellar mass, dark matter profile). We then describe how we build a stellar orbit library and construct a cold gas ring separately from this potential. Finally, we use the combined weights of stellar and cold gas kinematics to select our best-fitting model.

\subsection{Gravitational Potential}
\subsubsection{Stellar Mass}
\label{subsec:MGE}
The stellar mass distribution of a galaxy equals its surface brightness $S$ multiplied by its stellar mass-to-light ratio $\Upsilon$. 
We use the 2-dimensional Multiple Gaussian Expansion (MGE) \citep{emsellem1994multi,cappellari2002efficient} to describe the surface brightness, whose gravitational potential is analytical. This method makes use of Gaussian profiles to fit the total surface brightness profile of the galaxy. The surface brightness is described with the following equation: 
\begin{equation}
S(x',y') = \sum_{i} S_i(x',y') =\sum_{i} \left(\frac{L_i}{2\pi \sigma_i^2 {q_i}}\exp\left[-\frac{1}{2\sigma_i^2}\left(x'^2+\frac{y'^2}{q_i^2}\right)\right]\right),
\end{equation}
where $(x',y')$ are the two-dimensional coordinates aligned along the major and minor axis of the surface brightness profile; $S(x',y')$ is the surface brightness; $S_i(x',y')$ is the surface brightness distribution of each Gaussian component with corresponding amplitude $L_i$, scale length $\sigma_i$ and flattening $q_i$.

We apply MGE to the $r$-band Pan-STARRS image of NGC 2974. The resulting MGE model contains 7 Gaussians as shown in Table~\ref{tab:mge}. The residuals are within $2\%$ in the inner region and about $10\%$ in the outskirts of the galaxy. 
\begin{table}
  \centering
  \begin{tabular}{lllll}
  \hline
  $i$ & $L_i\mathrm{(L_\odot/pc^2)}$ & $\sigma$(arcsec) & $q_i$ & $M_i \mathrm{(M_\odot/pc^2)}$\\ \hline
       1   &      4276.01   &      0.54153   &     0.83144  &      16208.47 \\
       2   &      7782.37   &      0.88097   &     0.82501  &      26366.23 \\
       3   &      2853.55   &      1.44526   &     0.94271  &      13148.71 \\
       4   &      3171.34   &      3.81993   &     0.67267  &      11329.50  \\
       5   &      220.000   &      6.64704   &     0.99990  &       1966.17 \\
       6   &      970.160   &      10.7437   &     0.55375  &       2890.09 \\
       7   &      252.150   &      28.4453   &     0.61238  &        778.71 \\ \hline
  \end{tabular}
  \caption{MGE Parameters of the surface brightness and stellar mass of NGC 2974. From left to right: index, central luminosity intensity, width (standard deviation), axis ratio, central mass density of each Gaussian. The values of central mass density $M_i$ are already rescaled to the stellar mass-to-light ratio according to the Chabrier IMF.}
  \label{tab:mge}
\end{table}

To approximate the stellar mass also with an MGE model, based on the surface brightness, we introduce a group of free parameters $\tilde{\Upsilon_i}$,
\begin{equation}
M(x',y') = \sum_{i} \left[S_i(x',y') \cdot \tilde{\Upsilon_i}\right].
\end{equation}
Here $M(x',y')$ is the stellar mass distribution, and each $\tilde{\Upsilon_i}$ is a proxy for the mass-to-light ratio of the corresponding Gaussian component, albeit with no attached physical meaning.
Then the mass-to-light ratio $\Upsilon$ is defined as,
\begin{equation}
\Upsilon(x',y') \equiv \frac{M(x',y')}{S(x',y')}=\frac{\sum_{i} \left[S_i(x',y') \cdot \tilde{\Upsilon_i} \right]}{\sum_{i} S_i(x',y')}.
\end{equation}
which is a function of a group of unknown parameters $\tilde{\Upsilon_i}$ and the MGE parameters of the surface brightness. 

\begin{figure}
	\centering
	\includegraphics[width=0.9\columnwidth]{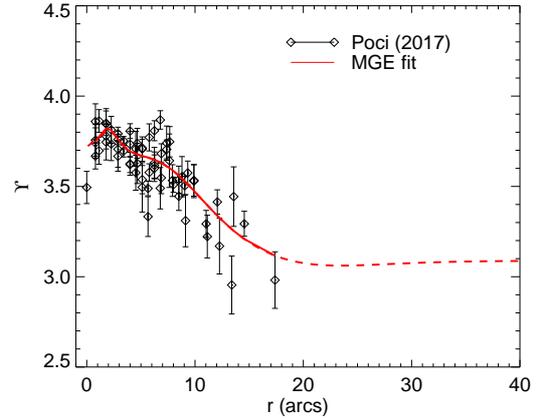}
    \caption{The $r$-band mass-to-light ratio of NGC 2974. The diamonds are data from \citet{poci2017systematic} but rescaled to Chabrier IMF; the solid line is the best fitting $\Upsilon$; the dashed line is a constant extension of $\Upsilon$ because of the limited radial coverage in \citet{poci2017systematic}.}
    \label{fig:ml_gradient}
\end{figure}

If $\Upsilon(x',y')=\Upsilon_0$ is a constant in the galaxy, all the $\tilde{\Upsilon_i}$ have the same value of $\Upsilon_0$. However, $\Upsilon(x',y')$ usually is not a constant and we need to decide $\tilde{\Upsilon_i}$ by fitting it. We choose the SDSS $r$-band mass-to-light ratio distribution \citep[][$\Upsilon$-maps available on Atlas3D website]{poci2017systematic}
as our $\Upsilon(x',y')$, which is compatible with the Pan-STARRS $r$-band image because of their comparable $r$-band filters. The resulting 1-dimensional fit of $\Upsilon(x',y')$ is shown in Figure~\ref{fig:ml_gradient}, and the corresponding MGE central mass density $M_i = L_i\cdot \tilde{\Upsilon_i}$ is shown in Table~\ref{tab:mge}.
We notice that there is a dip in $\Upsilon$ at the galaxy centre in Figure~\ref{fig:ml_gradient}, which is caused by a single data point. There is an AGN \citep{maia2003seyfert} in the centre of NGC 2974, which could cause this dip. We therefore neglect this single data point in the fitting of $\Upsilon(x',y')$.

The stellar mass and consequently $\Upsilon$ are affected by the initial mass function (IMF): for example, a Chabrier IMF \citep{chabrier2003galactic} produces almost 40\% less stellar mass than a Salpeter IMF \citep{salpeter1955luminosity} with the same observables \citep{santini2012evolving}. Here we introduce a factor $\alpha$ to indicate the stellar mass variation caused by the choice of IMF as a free parameter in our Schwarzschild modelling. Thus, the stellar mass distribution used in the Schwarzschild modelling becomes, 
\begin{equation}
M_\mathrm{mod}(x',y') = \alpha \cdot M(x',y') = \alpha \cdot \sum_{i} \left[S_i(x',y') \cdot \tilde{\Upsilon_i}\right].
\end{equation}
The mass-to-light ratio distribution in \citet{poci2017systematic} is obtained with the Salpeter IMF. We assume the galaxy has a constant IMF, hence $\alpha$ is a constant as well. We rescale the mass-to-light ratio such that the Chabrier IMF corresponds to $\alpha = 1$, and the Salpeter IMF corresponds to $\alpha = 1.7$ \citep{speagle2014highly}.

We deproject the 2-dimensional mass distribution to 3-dimensional mass density following \citet{cappellari2002efficient,van2008triaxial}, and introduce the intermediate and minor axis ratio $p_i$ and $q_i$. Since NGC 2974 is nearly axisymmetric with $p_i \sim 1$, we adopt only one viewing angle, the inclination $\theta$ as a free parameter in our model, with minor triaxiality still allowed.   

As the total \HI mass of NGC 2974 is three orders smaller than the stellar mass \citep{weijmans2008shape}, its contribution to the gravitational potential is negligible and therefore ignored in our model.

\subsubsection{Dark Matter Mass}
We adopt a spherical generalized NFW (gNFW) dark matter halo \citep{navarro1996structure,zhao1996analytical} with a density profile of 
\begin{equation}
\rho_r = \frac{\rho_\mathrm{s}}{(r/r_\mathrm{s})^\gamma [1+(r/r_\mathrm{s})^\eta]^{(3-\gamma)/\eta}}.
\end{equation}
This halo model has four free parameters: $\rho_\mathrm{s}$ is the scale density, $r_\mathrm{s}$ is the scale radius, $\gamma$ is the inner slope, while $\eta$ controls the turning point. The outer slope of this profile becomes $-3$ for $r \gg r_s$. When $\gamma = 1$ and $\eta = 1$, the gNFW halo profile reduces to the NFW profile. For $\gamma = 0$, the halo model has a core in its centre. 

We avoid calculating the gravitational potential of the gNFW profile analytically by expanding the halo density profile to an MGE as well. As the halo is spherical, we use the one-dimensional MGE expansion of \citet{cappellari2002efficient}. We then generate the total gravitational potential from the combined MGEs of stellar and dark matter halo density.

\subsubsection{Black Hole}
Based on the $M_\mathrm{BH}-\sigma$ relation \citep[e.g.][]{tremaine2002slope}, we expect a black hole mass of $M_\mathrm{BH} = 2.5 \times 10^8 \mathrm{M_\odot}$ for NGC 2974 \citep[see also][]{krajnovic2005dynamical}, which gives a radius of influence of just 0.2 arcsec. As this is below the spatial resolution of the SAURON spectrograph (0.8 arcsec), we at first neglected the contribution of the black hole to the gravitational potential. However, we did find that without the inclusion of the black hole, we could not reconstruct the observed velocity dispersions of the central regions in our models. We therefore decided to add the central black hole to the potential regardless, and modelled it as a point source:

\begin{equation}
\Phi_\mathrm{c,\mathrm{BH}} = -\frac{GM_\mathrm{BH}}{\sqrt[]{r^2+r_\mathrm{soft}^2}}.
\end{equation}
Here $r_\mathrm{soft}$ is the softening length of the black hole. We set this length to $r_\mathrm{soft} = 10^{-3}$ pc.

\subsection{Model of Stellar Kinematics}

All orbits in a separable potential are analytical through three conserved integrals of motion: $E$ (energy), $I_2$ and $I_3$. There are four different types of orbits: three types of tube orbits (short axis tubes, outer and inner long axis tubes) and the box orbits. Even if $I_2$ and $I_3$ are not analytic, three types of tubes orbits still exist, and most box orbits turn to be boxlets.

We sample the initial conditions of the orbits with their energy $E$ and their starting point on the $(x,z)$ plane \citep{van2008triaxial}. Each $E$ is linked with a grid radius $r_i$ such that $E$ equals the potential at position $(x,y,z)= (r_i,0,0)$. We sample $r_i$ logarithmically. For each energy, we then sample the starting point $(x,z)$ from a linear open polar grid $(R,\theta)$ in between the location of the thin orbits and the equipotential of this energy. The number of sample points across three dimensions $n_E \times n_\theta \times n_R = 21 \times 10 \times 7$. We introduce $3$ ditherings in every dimension ($E, \theta$ and $R$) and create an orbit bundle of $3 \times 3 \times 3$ dithering orbits for each orbit in our libraries to smooth the model, and this results in 5670 orbits in total. More details of the orbits sampling can be found in \citet{van2008triaxial}. 

This orbit library includes mostly short and long axis tubes and hardly contains box orbits in the inner region. In practice however, early-type galaxies are not perfectly axisymmetric and should contain a number of box orbits. To generate enough box orbits for a triaxial shape, we also add an additional box orbit library dropped from the equipotential surface following the method in \citet{zhu2018orbital}, using energy $E$ and two spherical angles $\theta$ (inclination) and $\phi$ (azimuthal angle). Energy $E$ and inclination $\theta$ are sampled in the same way as the first library, while $\phi$ is linearly sampled. The number of sample points $n_E \times n_\theta \times n_\phi$ for this library also equals $21 \times 10 \times 7$. We also smooth the model by introducing $3$ ditherings in $E, \theta$ and $\phi$ for each orbit and create 5670 orbits in the box orbit library.

\subsection{Model of Cold Gas Kinematics}
We assume that the \HI gas is dynamically cold, and hence neglect its velocity dispersion. Since the typical \HI velocity dispersion in discs is less than 10 km/s, this is a reasonable assumption for our case, where typical velocity errors are of the same order.

We also assume that the \HI gas forms an axisymmetric thin ring aligned with the stellar disc in the equatorial plane of the galaxy. This again is a reasonable assumption for NGC 2974, given that the velocity fields of the \HI ring and the stellar discs are aligned (see Figure~\ref{fig:stellar} and~\ref{fig:hi}). The \HI gas moves on circular orbits on the disc plane with a velocity of
\begin{equation}
V_\mathrm{c} = \left. \sqrt[]{\frac{\partial \Phi}{\partial r}} \right|_{z=0},
\end{equation}
where $\Phi$ is the total gravitational potential, including stars, dark matter halo and black hole. As shown, the \HI velocity allows us to constrain the total gravitational potential although it provides no constraints on the stellar orbit distribution.

The model light-of-sight velocity is given by
\begin{equation}
v_\mathrm{mod} = V_\mathrm{c} \sin \theta \cos\phi,
\end{equation}
where $\phi$ is the azimuthal angle from the major axis and $\theta$ is the inclination. To compare the model and observational velocity directly, we need to convolve the model velocity to take the beam smearing into account. 
We adopt a homogeneous \HI mass distribution, therefore, we can directly convolve the model velocity map with the beam.

\subsection{Combining Kinematics Weights}
Our method requires two different data sets: the stellar kinematics (including the zero moment or surface brightness, as described by the MGE model), and the cold gas kinematics. The total $\chi^2$ for each model built from the model parameters therefore contains two terms:
\begin{equation}
\chi^2 = \chi_\mathrm{star}^2 + \chi_\mathrm{gas}^2.
\end{equation}
The best-fitting models are determined by selecting the models with the minimum $\chi^2$.

The stellar surface brightness and kinematics are reproduced simultaneously in a single Schwarzschild model by a superposition of all orbit bundles. Each orbit bundle $k$ has a weight $w_k$. The weights of these orbit bundles are solved by minimising $\chi_\mathrm{star}^2$ as:
\begin{equation}
\chi_\mathrm{star}^2 = \chi_\mathrm{s,lum}^2 + \chi_\mathrm{s,kin}^2.
\end{equation}
In practise we only take residuals of stellar kinematics $\chi_\mathrm{s,kin}^2$ into consideration, because the residual of the surface brightness distribution fitting $\chi_\mathrm{lum}^2$ is negligible compared to the other terms. Once the orbit bundle weights of a model are solved, the $\chi_\mathrm{s,kin}^2$ for this model is fixed.

The model confidence level for \emph{all} Schwarzschild models is enlarged by the fluctuation of $\chi_\mathrm{s,kin}^2$, which has a standard deviation of $\sim \sqrt{2N_\mathrm{s,kin}}$, where $N_\mathrm{s,kin} = 6924$ is the total number of stellar kinematic data. We therefore set $\mathrm{\Delta}\chi_\mathrm{star}^2 = \sqrt{2N_\mathrm{s,kin}}$ as the $1-\sigma$ confidence level. A more detailed description of this method for calculating $\chi_\mathrm{star}^2$ can be found in \citet{van2008triaxial,van2009recovering,zhu2018orbital}. 

The residual of cold gas kinematics for each model $\chi_\mathrm{gas}^2$ equals 
\begin{equation}
\chi_\mathrm{gas}^2 = \sum \left(\frac{v_\mathrm{mod}-v_\mathrm{los}}{\epsilon_\mathrm{los}}\right)^2,
\end{equation}
with $\epsilon_\mathrm{los}$ the error in observed velocity. Our total number of gas kinematic data $N_\mathrm{g,kin} = 1732$ and $\sqrt{2N_\mathrm{g,kin}} = 59$. 
However, the $1-\sigma$ confidence level for fitting the cold gas kinematics is larger than $\sqrt{2N_\mathrm{g,kin}}$. When we perturb the \HI velocity by adding random Gaussian noise to the kinematic data with the standard deviations of the Gaussian noise being the $1-\sigma$ uncertainties of \HI velocity, $\chi_\mathrm{gas}^2$ fluctuates strongly with a standard deviation of $\mathrm{\Delta}\chi_\mathrm{gas}^2 = 310$. Therefore, we set this value as the $1-\sigma$ confidence level for fitting the cold gas kinematics.

Combining the confidence intervals for stellar and cold gas data, the $1-\sigma$ confidence level for \emph{all} models is
\begin{equation}
\mathrm{\Delta}\chi_\mathrm{tot}^2 =  \mathrm{\Delta}\chi_\mathrm{star}^2 + \mathrm{\Delta}\chi_\mathrm{gas}^2.
\end{equation}

\section{Results}
We apply our modelling technique to NGC 2974 as described in Section~\ref{sec:method}. In total we generated $4,259$ dynamical models of NGC 2974, and these models are selected in two different ways: by including both the cold gas kinematics and stellar kinematics, and by fitting the stellar kinematics only.

\subsection{Parameter Grid}
We have 5 free parameters in total: these are the IMF factor $\alpha$, the three parameters of the dark matter halo profile ($\rho_\mathrm{s}$,$r_\mathrm{s}$,$\gamma$) and the black hole mass $M_\mathrm{BH}$. 
As \citet{weijmans2008shape} have already shown that the \HI ring has an inclination of $60\pm 2\degree$, we fix the inclination $\theta$ to $60 \degree$. The turning point of the dark matter profile $\eta$ is fixed to $2$ in our model because we find that $\eta$ is not well constrained even with cold gas kinematics. 

The IMF factor $\alpha$ varies from $1.0$ to $2.0$ in steps of $0.1$, to represent different IMFs.
For the dark matter profile, the central density $\rho_\mathrm{s}$ and the scale radius $r_\mathrm{s}$ are sampled on a logarithmic grid, $\log[\rho_\mathrm{s}/(\mathrm{M_\odot}f\cdot\mathrm{pc}^{-3})] \in [-5,1]$ and $\log(r_\mathrm{s}/\mathrm{pc}) \in [3,5]$. The inner slope of the dark matter profile,  $\gamma \in [0,1]$ is in a linear grid with a step of $0.1$, and has a minimum step of $0.05$ around the best-fitting model. 
The black hole mass $M_\mathrm{BH}$ is sampled on a logarithmic grid, with $\log(M_\mathrm{BH}/\mathrm{M_\odot})$ sampled on the interval $[6, 10]$ in steps of $0.25$.

\begin{figure*}
	\centering
	\includegraphics[width=0.7\paperwidth]{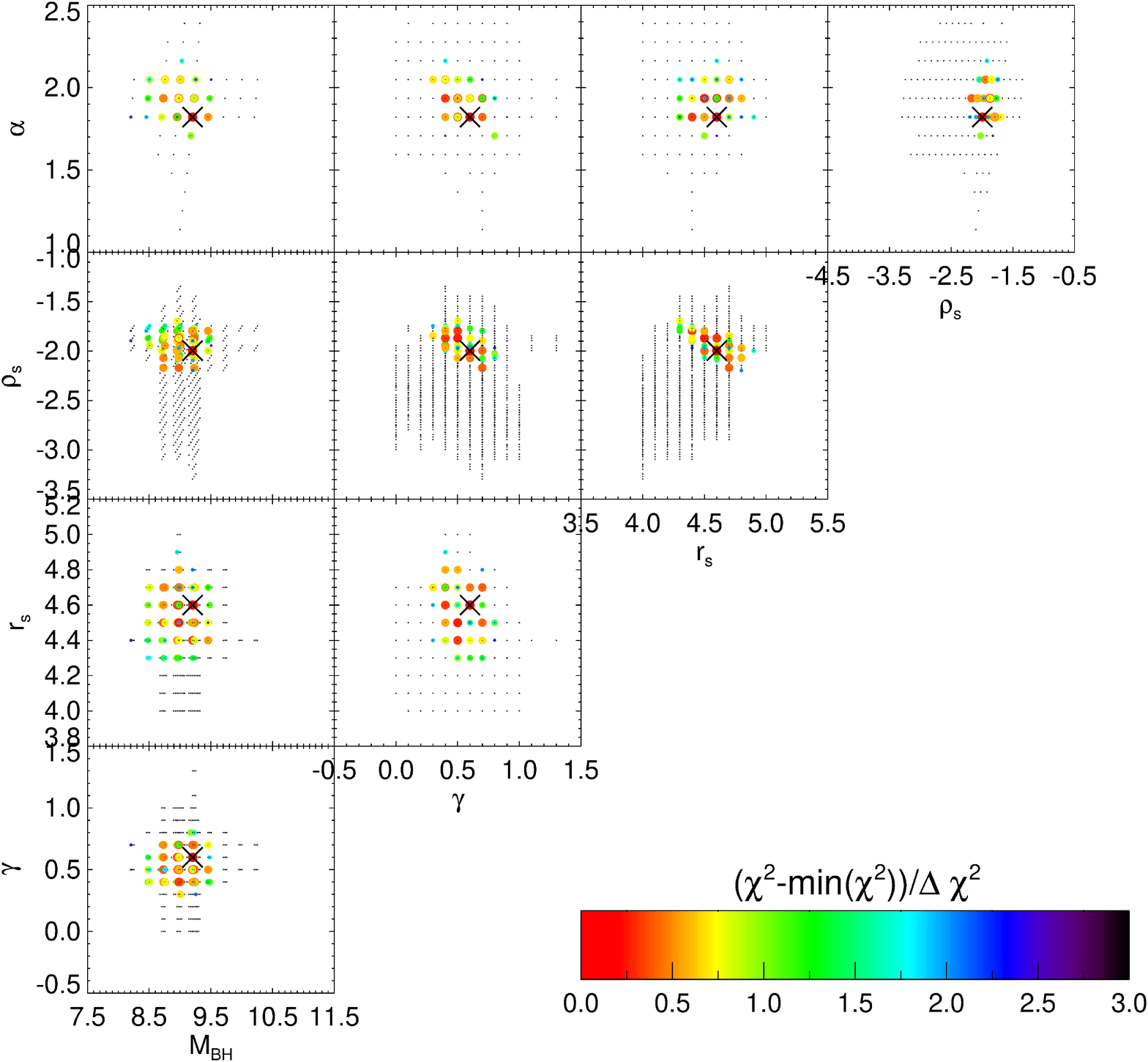}
    \caption{The grids of parameter space with cold gas constraints. The best-fitting model is marked with a cross sign. The coloured dots represent models within $3-\sigma$ confidence level, and larger and redder dots stand for models with smaller $\chi^2_\mathrm{tot}$. The small black dots are the remaining models. }
    \label{fig:chi2_grid}
\end{figure*}

\begin{figure*}
	\centering
    \includegraphics[width=0.7\paperwidth]{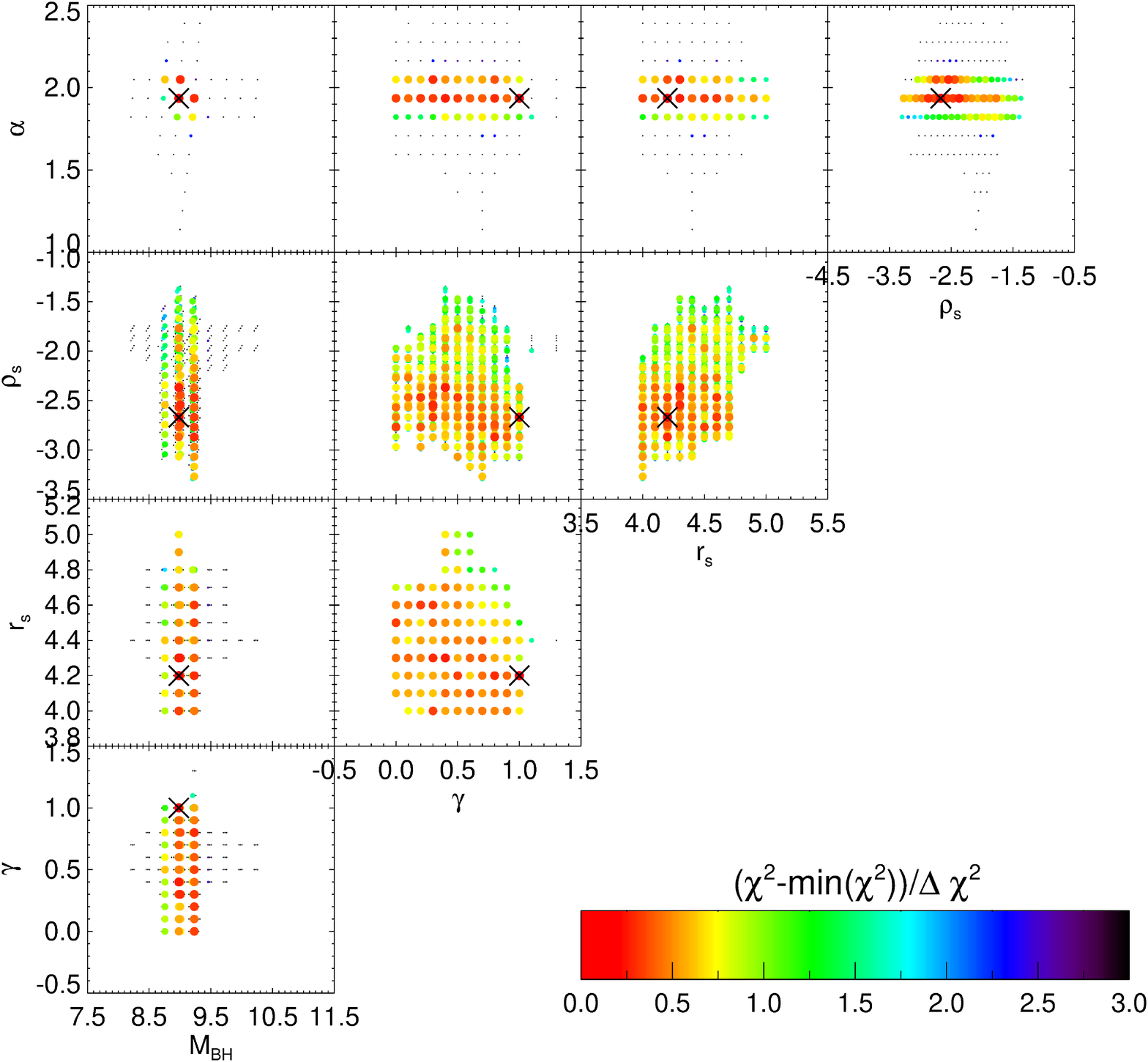}
    \caption{The grids of parameter space without cold gas constraints. The best-fitting model is marked with a cross sign. The coloured dots represent models within $3-\sigma$ confidence level, and larger and redder dots stand for models with smaller $\chi^2_\mathrm{star}$. The small black dots are the remaining models. }
    \label{fig:chi2_grid_nogas}
\end{figure*}

We plot the parameter grids as 2D projections of the parameter space shown in Figure~\ref{fig:chi2_grid} and~\ref{fig:chi2_grid_nogas}. 
The dots represent all the models we have run, and the coloured dots are the models within $3-\sigma$ confidence level, where $1-\sigma$ confidence levels are defined as $\chi^2_\mathrm{tot} - \mathrm{min}(\chi^2_\mathrm{tot}) <  \mathrm{\Delta}\chi^2_\mathrm{tot}$,  and $\chi^2_\mathrm{star} - \mathrm{min}(\chi^2_\mathrm{star}) <  \mathrm{\Delta}\chi^2_\mathrm{star}$ for the cases with and without gas, respectively.
It is obvious that cold gas constraints significantly lessen the models within $3-\sigma$ confidence level and reduce the uncertainties of the fitting, especially for parameters related to the dark matter profile $\rho_\mathrm{s}$, $r_\mathrm{s}$ and $\gamma$. 

When including the cold gas kinematics in the fit, $48$ of the $4,259$ models fall within the $1-\sigma$ confidence interval, while this number increases to $1110$ models if the cold gas kinematics are omitted. This already demonstrates the value of including the constraints of cold gas data, as the parameters of the models get more tightly constrained.

\subsection{Best-fitting Model}
The best-fitting parameters for our model are shown in Table~\ref{tab:best}. The best-fitting parameters are the values of the best-fitting model identified as having the smallest $\chi^2$, and the uncertainties quoted as the lower and upper limits of all models within the $1-\sigma$ confidence level. We list both the best-fitting parameter for the models with and without cold gas constraints. The parameters of the dark matter profile are better constrained by including the cold gas measurements, even though their actual values do not change significantly between the two different models.
\begin{table*}
\renewcommand\arraystretch{1.5}
\centering
\begin{tabular}{llllllllll}
\hline
 & $\alpha$ & $\rho_\mathrm{s}(10^{-3} \mathrm{M_\odot}/\mathrm{pc^3})$ & $r_\mathrm{s}$(kpc) & $\gamma$ & $M_\mathrm{BH} (10^9 \mathrm{M_\odot})$ & $r_{200}\mathrm{(kpc)} $ & $M_{200}(10^{13}\mathrm{M_\odot})$ & $c$ \\ \hline
star + gas &  $1.8^{+0.2}_{-0.1}$ & $10.1^{+10.0}_{-3.3} $ & $40^{+23}_{-20}$  & $0.6^{+0.2}_{-0.3}$ &  $1.6^{+1.2}_{-1.3}$ & $560^{+340}_{-200}$ & $2.0^{+6.3}_{-1.5}$  & $14^{+4}_{-2}$  \\
star only &  $1.9^{+0.1}_{-0.1}$ & $2.1^{+31.8}_{-1.6} $ & $16^{+84}_{-6}$  & $1.0^{+0.0}_{-1.0}$ &  $1.0^{+0.7}_{-0.4}$ & $130^{+1430}_{-80}$ & $0.022^{+43.125}_{-0.021}$  & $8^{+14}_{-4}$ \\\hline
\end{tabular}
\caption{The best-fitting parameters and deduced dark matter halo parameters (virial mass $M_{200}$, virial radius $r_{200}$ and concentration $c$, see Section~\ref{subsec:dmp}) for our two fitting cases: with and without cold gas constraints. The uncertainties are the lower and upper limits of all models within $1-\sigma$ confidence level. Here $\rho_\mathrm{s}$ and $M_\mathrm{BH}$ are already multiplied by $\alpha$ to obtain their actual values.}
\label{tab:best}
\end{table*}

$\alpha$ is around 1.8, and it produces a total stellar mass $6\%$ more than the total mass produced by the Salpeter IMF in \citet{poci2017systematic}. We are however not able to infer the shape of IMF from $\alpha$.
The stellar $\Upsilon$ in $r$-band is $5.4 \mathrm{M_\odot/L_\odot}$ in the outskirts and rises up to $6.8 \mathrm{M_\odot/L_\odot}$ in the centre. 
\citet{cappellari2013atlas3d2} measure a stellar $\Upsilon$ in $r$-band of $8.9 \mathrm{M_\odot/L_\odot}$ with the assumption of a NFW halo, which is significantly heavier than our measurements. However, when they assume a Salpeter IMF, their $\Upsilon$ within $1 R_\mathrm{e}$ is $6.1 \mathrm{M_\odot/L_\odot}$, consistent with our measurements.

\citet{krajnovic2005dynamical} include a central black hole in their Schwarzschild model of NGC 2974 with a fixed mass as predicted by the $M_\mathrm{BH}-\sigma$ relation ($2.5\times 10^8 \mathrm{M_\odot}$)\citep[e.g.][]{tremaine2002slope}. Our best-fit models return a black hole mass of $1.6^{+1.2}_{-1.3} \times 10^9 \mathrm{M_\odot}$. However, as we stated before, due to the spatial resolution of our data, as well as the not well understood behaviour of $\Upsilon$ in the very central parts of the galaxy, we do not make any claims about the true mass of the supermassive black hole in NGC 2974 from our models.

\begin{figure*}
	\centering
	\includegraphics[width=0.7\paperwidth]{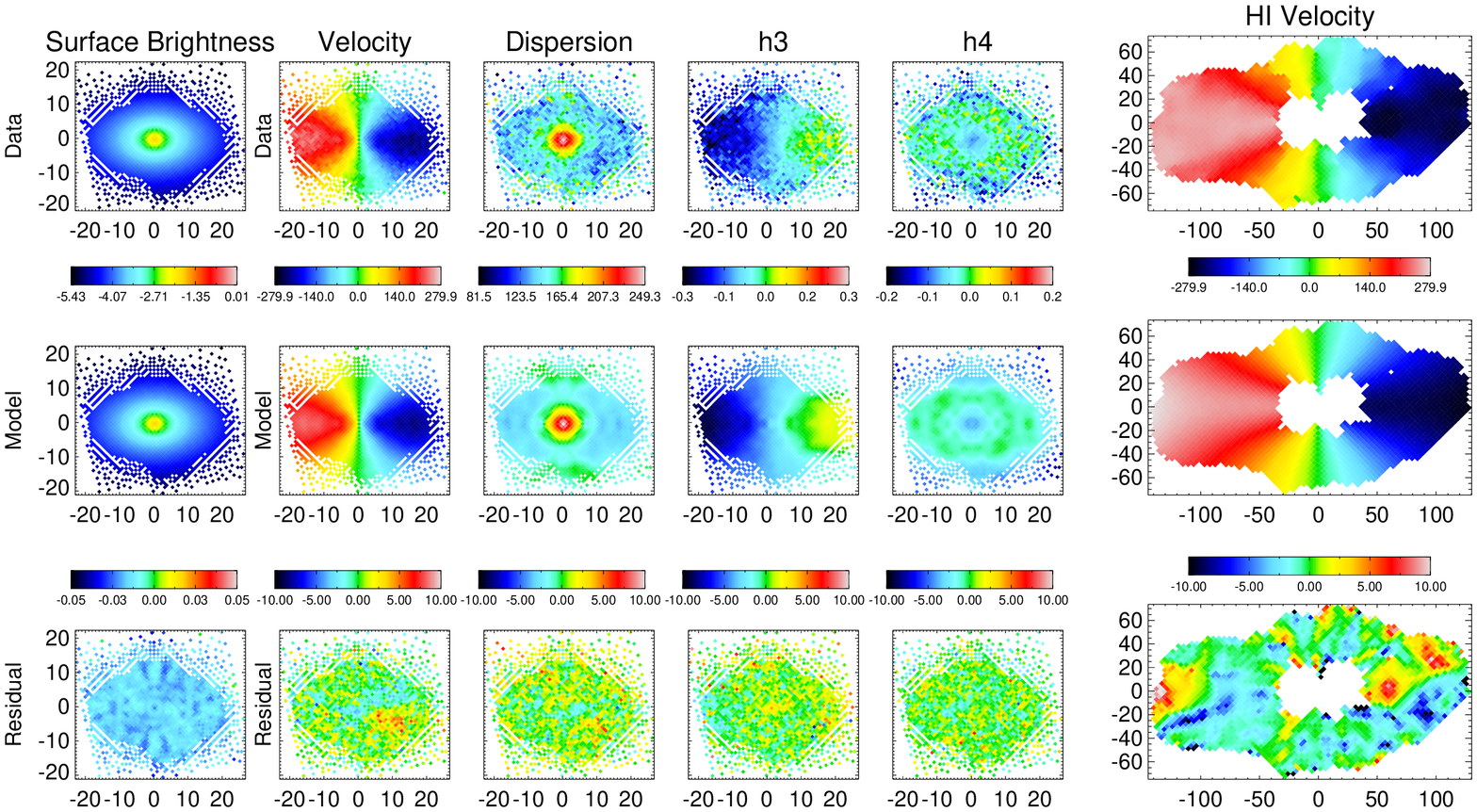}
    \caption{The data (top), model (middle) and relative residual (bottom; defined as (data-model)/error) of the surface brightness, stellar velocity, velocity dispersion, the third and fourth orders of Gauss-Hermite moments and cold gas velocity of NGC 2974 from left to right. }
    \label{fig:kin}
\end{figure*}

The corresponding best-fitting surface brightness and kinematics maps for the model with cold gas constraints are shown in Figure~\ref{fig:kin}. The residual plots do not show strong sub-structures.

\subsection{Dark Matter Profile}
\label{subsec:dmp}
Two parameters are commonly used to describe the dark matter profile: the virial mass $M_{200}$, which is defined as the enclosed mass within the virial radius $r_{200}$, where the average density within $r_{200}$ is 200 times the critical density ($\rho_\mathrm{crit} = 1.37\times 10^{-7} \mathrm{M_\odot/pc^3}$, adopting a Hubble constant $H_0 = 70 \mathrm{km/s/Mpc}$); and the concentration $c$, which is defined as the ratio of the viral radius $r_{200}$ and the scale radius $r_\mathrm{s}$. These parameters are listed in Table~\ref{tab:best} for our best-fitting models, which shows that the dark matter profile is better constrained by including the cold gas kinematics in the fit.

\subsubsection{Enclosed Mass Profile}
\begin{figure*}
	\centering
	\subfigure[cold gas included]{\includegraphics[width=0.9\columnwidth]{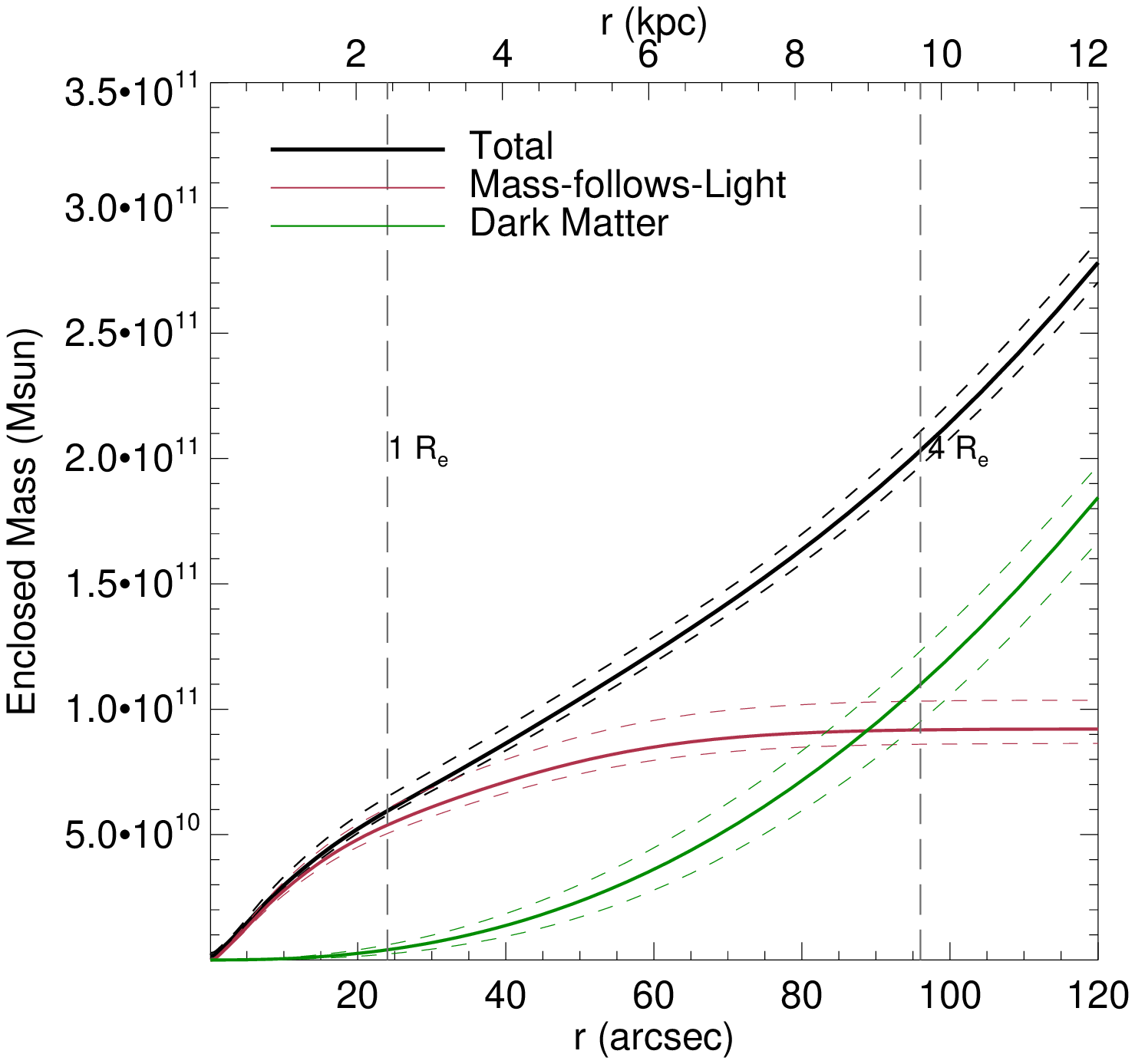}}
    \subfigure[cold gas excluded]{\includegraphics[width=0.9\columnwidth]{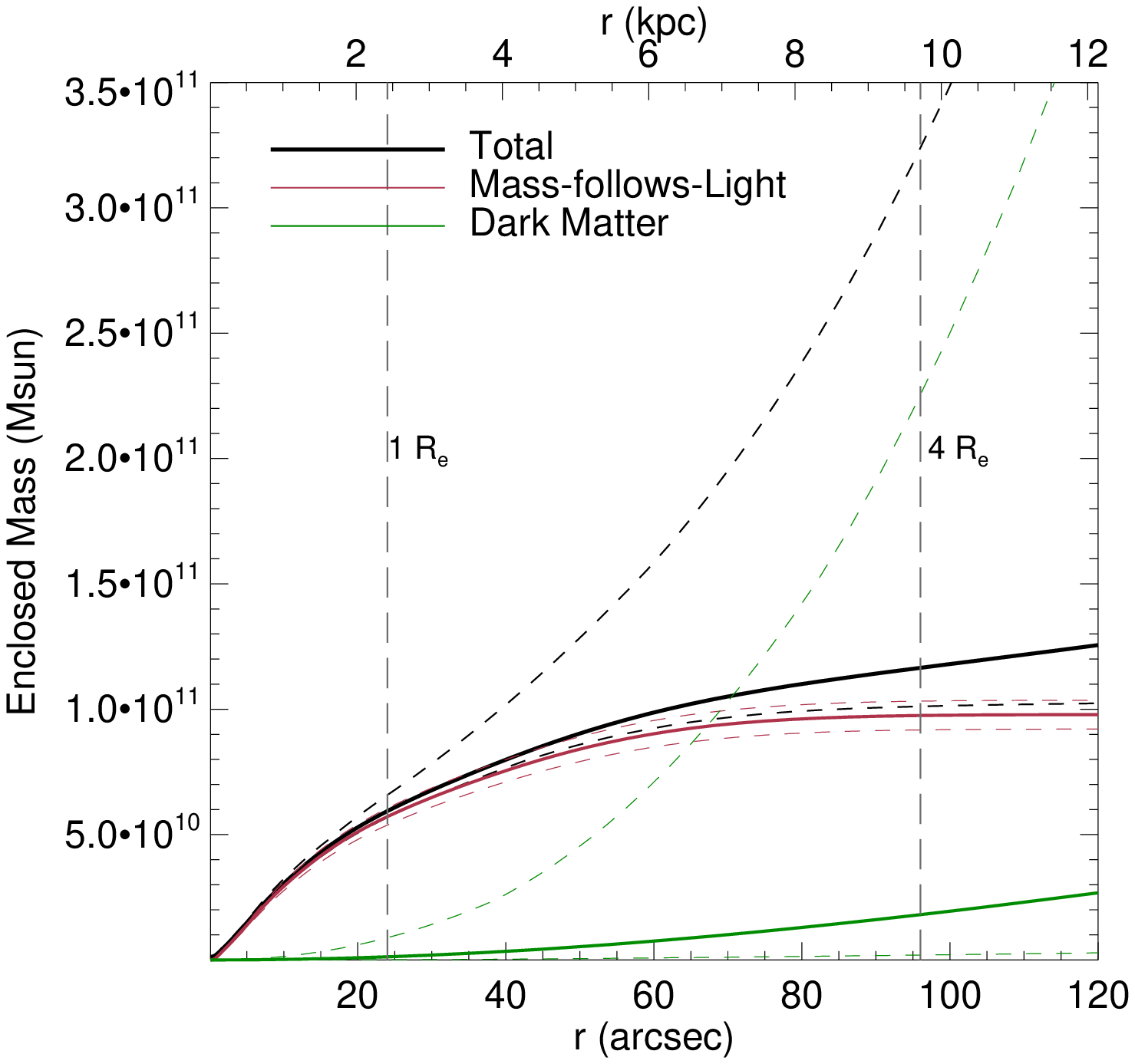}}
    \caption{The enclosed mass profile of NGC 2974: (a) modelling with cold gas constraints, (b) modelling without cold gas. The black, red and green solid lines stand for the total, stellar and dark matter mass, respectively. The corresponding dashed lines are their $1-\sigma$ uncertainties. The dark matter fraction is measured with much smaller uncertainty for the model that includes the cold gas constraints. The red dashed line representing the lower uncertainty overlaps with the red solid line in the right-hand panel.}
    \label{fig:enclosed}
\end{figure*}

The enclosed mass profiles of NGC 2974 are shown in Figure~\ref{fig:enclosed}, both for the cases with and without cold gas constraints. 
The black, red and green solid lines stand for the enclosed total, stellar and dark matter mass, respectively. The corresponding dashed lines show the lower and upper limits for the models within $1-\sigma$ confidence level. 
The stellar mass profile changes little between the two plots, as it dominates the inner region within $2 R_\mathrm{e}$ and is mainly constrained by the stellar kinematic data. The dark matter fraction is $7\%$ within $1 R_\mathrm{e}$, consistent with the measurement in \citet{cappellari2013atlas3d2} and smaller than the measurements of \citet{weijmans2008shape} and \citet{poci2017systematic}.
The dark matter fraction is $66\%$ within $5 R_\mathrm{e}$ with an uncertainty of $10\%$. It is significantly better constrained with the cold gas covering the region outside $4 R_\mathrm{e}$, where the dark matter begins to be dominating. This emphasizes the importance of extended tracers for dark matter measurements.

\subsubsection{Dark Matter Inner Slope}
\begin{figure*}
	\centering
    \subfigure[cold gas included]{\includegraphics[width=0.9\columnwidth]{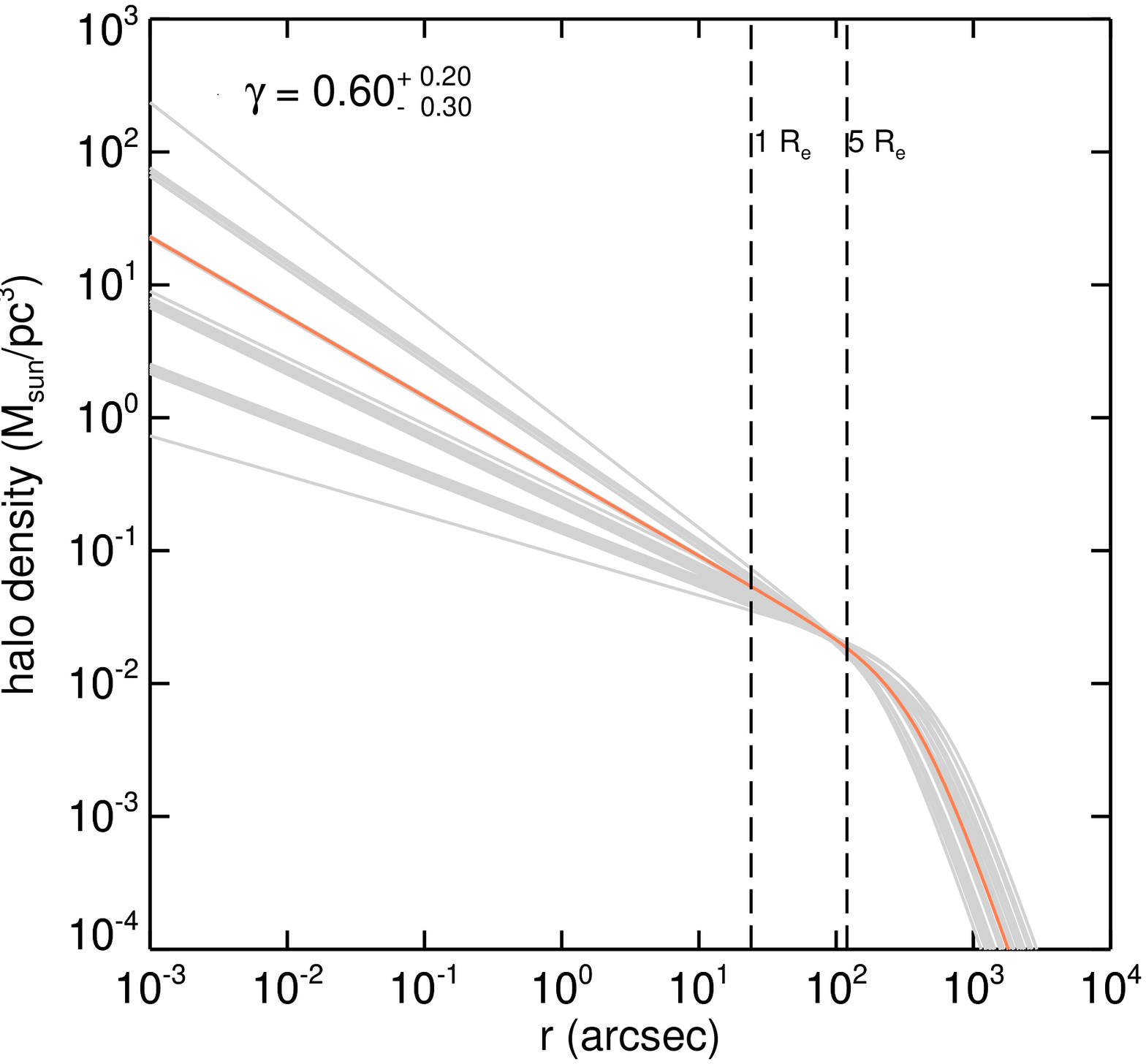}}
	\subfigure[cold gas excluded]{\includegraphics[width=0.9\columnwidth]{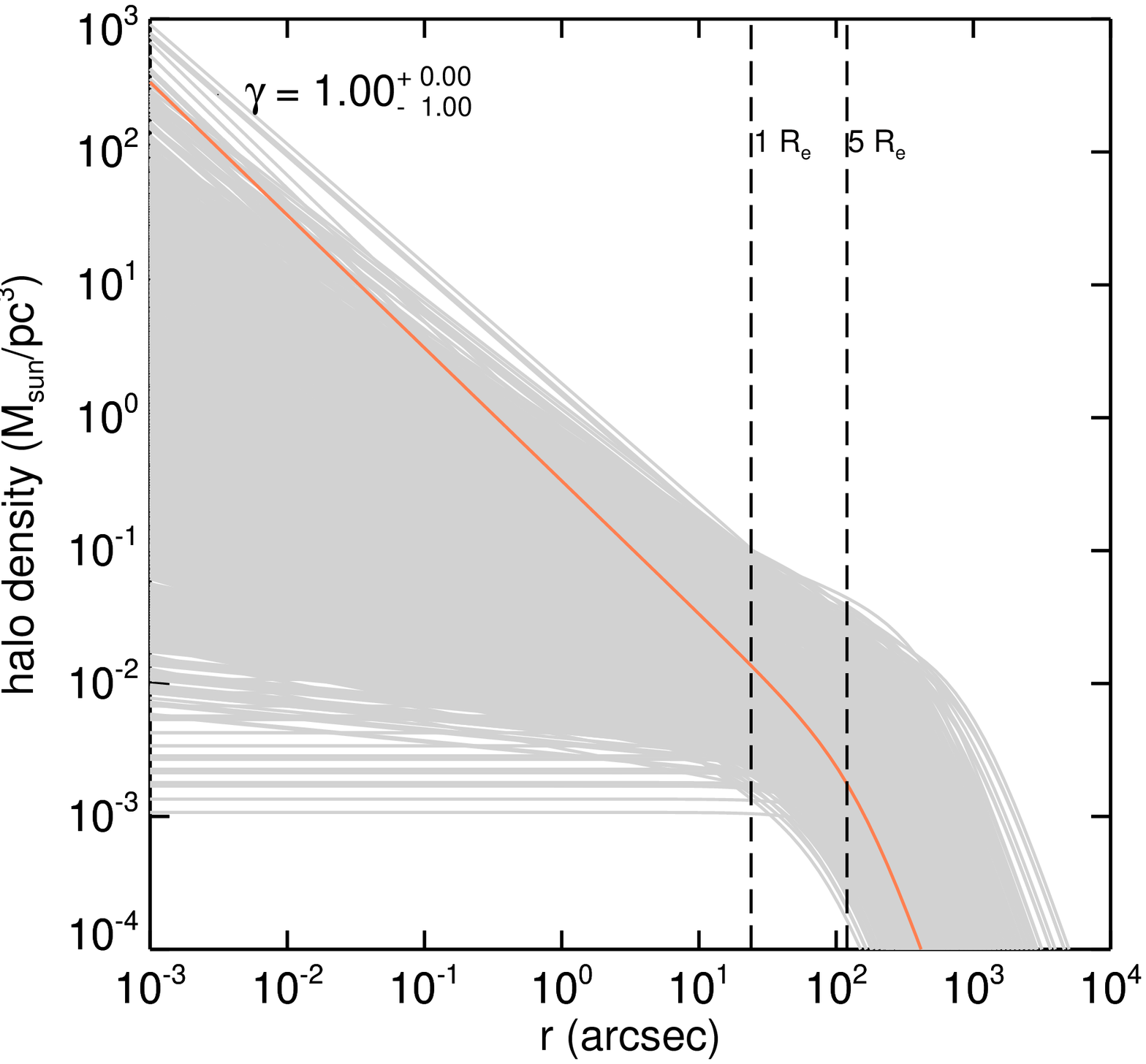}}
    \caption{The dark matter profiles of all models within $1-\sigma$ uncertainties: (a) modelling with cold gas constraints, (b) modelling without cold gas. Each grey line represents the dark matter profile of a model, and the orange line is the profile of the best-fitting model. We also list the inner slope $\gamma$ of the best-fitting model and its uncertainty. The dark matter halo inner slope is much better constrained by including the cold gas kinematics.}
    \label{fig:halo}
\end{figure*}

We show the dark matter density profiles for the models within $1-\sigma$ confidence level for both cases with and without the cold gas constraints in Figure~\ref{fig:halo}. It is apparent that the dark matter profile is constrained much better by including the cold gas kinematics. The inner slope $\gamma$ of NGC 2974 seems to prefer a shallow cuspy profiles.

Figure~\ref{fig:halo} again highlights the significance of the cold gas constraints. Including the cold gas kinematics in the model significantly reduces the uncertainties of $\gamma$.
As shown in the right panel of this figure, with the stellar kinematics only the $1-\sigma$ uncertainties include the full parameter range for $\gamma$, indicating we cannot constrain $\gamma$ without the cold gas kinematics.

\subsection{Stellar Orbit Distribution}

\begin{figure*}
	\centering
	\subfigure[cold gas included]{\includegraphics[width=0.9\columnwidth]{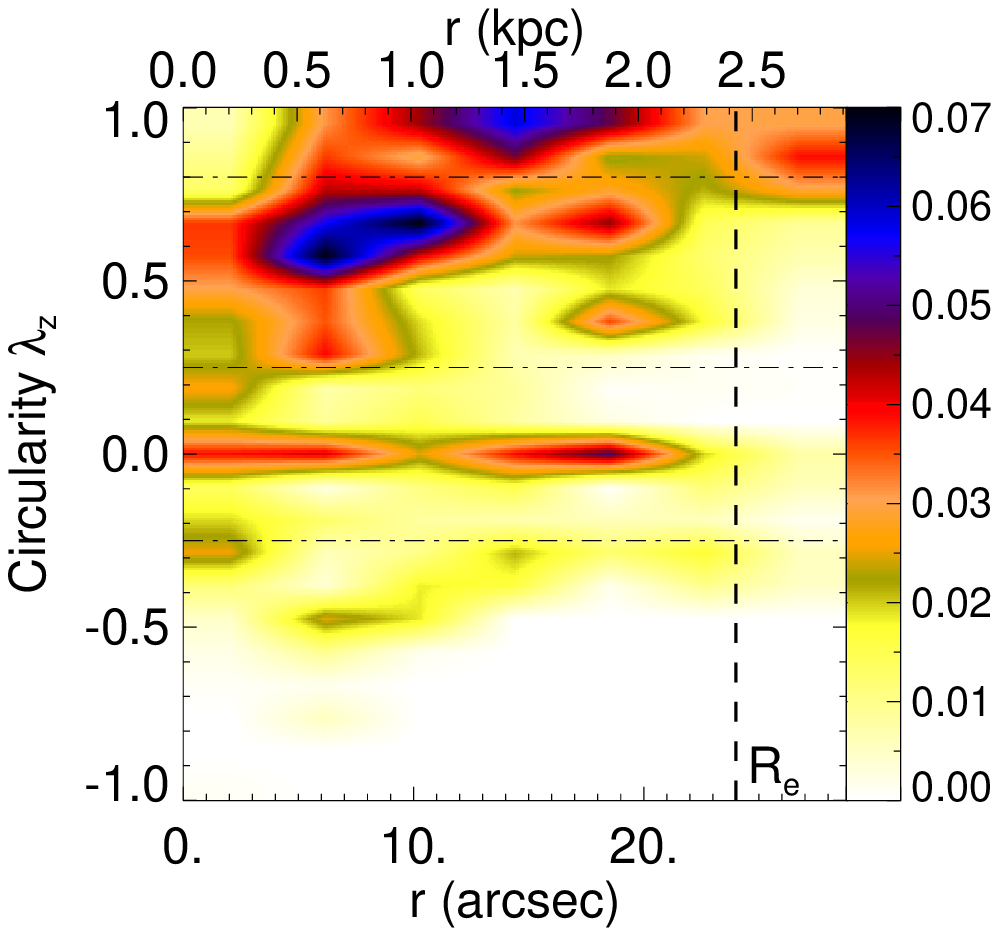}}
    \subfigure[cold gas excluded]{\includegraphics[width=0.9\columnwidth]{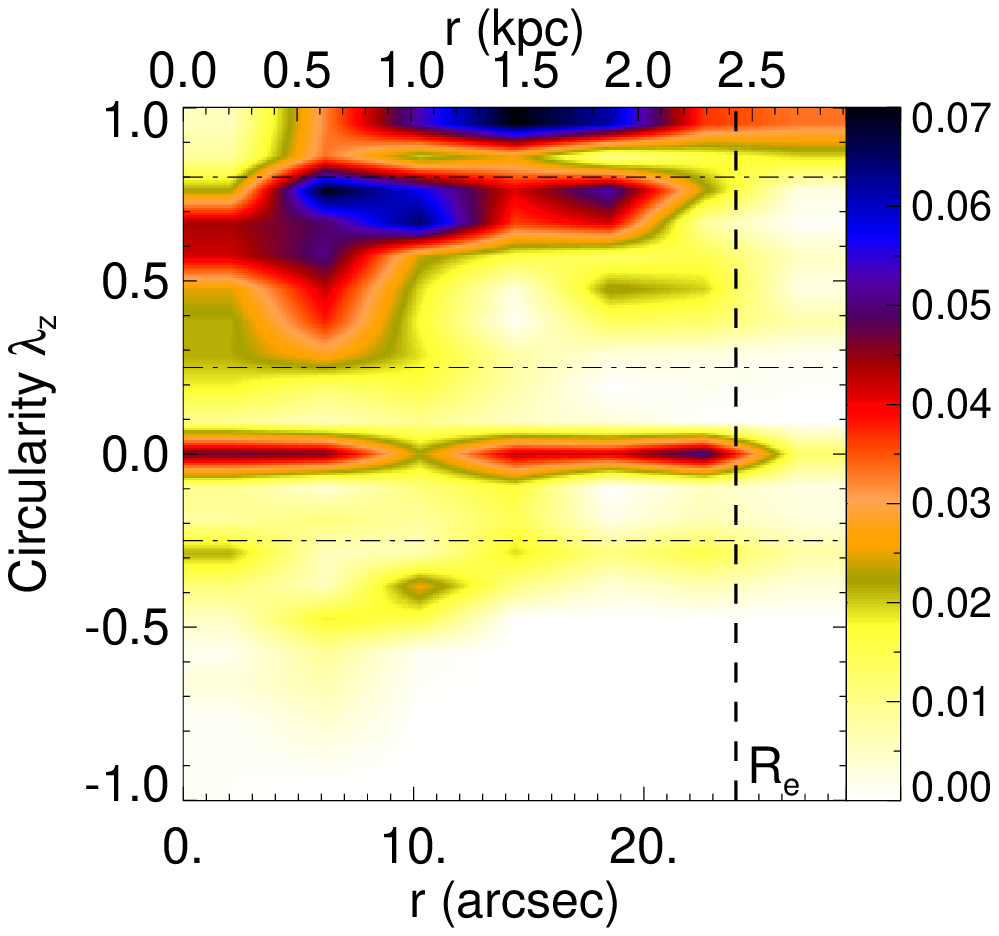}}
    \caption{The stellar orbit distribution on the phase-space of $\lambda_z$ vs. $r$ of the best-fitting model to NGC 2974: (a) modelling with cold gas constraints, (b) modelling without cold gas. The color bar indicates the probability density of orbits. In both cases, we discern a hot central component, an extended warm central component, and an extended cold component.}
    \label{fig:Lz_orbit}
\end{figure*}

The stellar orbit distribution offers information on the galaxy components and morphology, based on their model kinematics. We characterize the stellar orbits with their circularity, defined as the ratio of circular motion and total motion as:
\begin{equation}
\lambda_z = \overline{L_z}/(r\overline{V_c}),
\end{equation}
where $\overline{L_z} = \overline{xv_y-yv_x}$, $r = \overline{\sqrt{x^2+y^2+z^2}}$ and $\overline{V_c} = \sqrt{\overline{v_x^2+v_y^2+v_z^2+2v_xv_y+2v_xv_z+2v_yv_z}}$, taken the average for each orbit.
Based on this parameter, we classify the orbits in our model into three dynamical components: hot ($\lambda < 0.25$), warm ($0.25 < \lambda < 0.8$), and cold ($\lambda > 0.8$). 
The circularity map (Figure~\ref{fig:Lz_orbit}) shows that we can distinguish three major dynamical components in our dynamical model of NGC 2974: an extended hot component related to a prominent bulge, a central warm component possibly representing a thick disc, and an extended cold component linked to a thin disc. This is consistent with  NGC 2974 being a lenticular galaxy. 
The cold gas constraints lead to little differences in the stellar orbit distribution in the best-fitting model. 
Compared to the axisymmetric orbit-superposition model made by \citet{krajnovic2005dynamical}, our method produces a strong hot component instead of a strong counter-rotating component, because we allow a triaxial shape and sample enough box orbits by adding an additional box library, while an axisymmetric model cannot generate box orbits at all.

\section{Discussion}
One of our goals was to constrain the inner slope $\gamma$ of the gNFW dark matter profile of NGC 2974. A wide range of values is quoted in literature for $\gamma$ in early-type galaxies, e.g. \citet{wasserman2018sluggs} modelled the massive elliptical galaxy NGC 1407 and found an NFW-like inner slope $\gamma = 1.0_{-0.4}^{+0.2}$, consistent with the average inner slope $\gamma = 0.80_{-0.22}^{+0.18}$ of 81 strong lenses early-type galaxies  as obtained by \citet{sonnenfeld2015sl2s}. 
However, inner slopes steeper than NFW expectation are also reported: \citet{grillo2012average} showed that the average logarithmic slope of 39 strongly lensed early-type galaxies is $\gamma = 2.0 \pm 0.2$ or $\gamma = 1.7 \pm 0.5$ with a Chabrier or Salpeter-like IMF, respectively. 
\citet{mitzkus2016dominant} also found $\gamma = 1.4 \pm 0.3$ in a gNFW dark matter profile for lenticular galaxy NGC 5102.
\citet{oldham2018dark} obtained similar results, showing that the majority of massive early-type galaxies have an average $\gamma = 2.09_{-0.22}^{+0.19}$.
Yet a core-like dark matter profile is not ruled out for early-type galaxies: \citet{forestell2010hobby} found a power-law slope of $0.1$ in their best-fit dark halo model and rule out the NFW profile at $99\%$ confidence level. \citet{zhu2016discrete} also found that a core model is preferred for massive elliptical NGC 5846, although a cusp model would still be acceptable. A small number of early-type galaxies in \citet{oldham2018dark} are consistent with cored models (average $\gamma = 0.10_{-0.10}^{+0.33}$).
We find an inner slope $\gamma = 0.6^{+0.2}_{-0.3}$ in NGC 2974, consistent with a shallow cuspy profile.

\begin{figure}
	\centering
	\includegraphics[width=0.8\columnwidth]{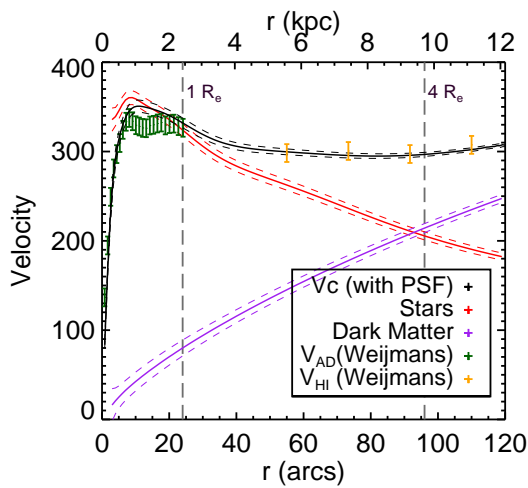}
    \caption{The circular velocity of NGC 2974. The black solid line is our rotation curve calculated from the gravitational potential and convolved with the SAURON PSF; the red and purple solid lines are the stellar and dark matter contributions without the PSF convolution; the corresponding dashed lines are their $1-\sigma$ uncertainties. The error-bars are from \citet{weijmans2008shape}: the green error-bars are the stellar velocity after asymmetric drift correction and the orange error-bars are the \HI velocity. The difference at 1 kpc is because \citet{weijmans2008shape} adopt an analytical model based on asymmetric drift correction.}
    \label{fig:ad}
\end{figure}

\citet{weijmans2008shape} corrected observed stellar velocities of NGC 2974 for asymmetric drift, to obtain the circular velocity representative of the gravitational potential of the galaxy. In Figure~\ref{fig:ad} we show a comparison between their rotation curve, and the one we extracted from our Schwarzschild models.
The black solid and dashed lines are the rotation curve and corresponding uncertainties generated from the gravitational potential of our Schwarzschild models within the $1-\sigma$ confidence level. The rotation curve of \citet{weijmans2008shape} has two parts: the orange dots show the \HI velocity data that we also used for our models; the green dots are the stellar velocities corrected for asymmetric drift. Our model is therefore consistent with the earlier work by \citet{weijmans2008shape}. 

\section{Summary}
We introduced an orbit-based method with combined stellar and cold gas kinematics and applied it to early-type galaxy NGC 2974. The main results are as following.

\begin{enumerate}
\item Our modelling shows a preference for a shallow cuspy dark matter halo profile, with the inner slope of the halo $\gamma =0.6^{+0.2}_{-0.3}$ in a gNFW profile. The dark matter halo has a total mass of $M_{200}= 2.0^{+6.3}_{-1.5} \times 10^{13}\mathrm{M_\odot}$ and a concentration of $c = 10.8^{+3.2}_{-1.5}$. 
We also find that the stellar mass is slightly heavier than the mass produced if we assume a Salpeter IMF, with a corresponding stellar $\Upsilon$ in $r$-band decreasing from $6.8 \mathrm{M_\odot/L_\odot}$ in the centre to $5.4 \mathrm{M_\odot/L_\odot}$ in the outskirts.

\item The comparison between the results of the Schwarzschild modelling with and without cold gas clearly shows that the cold gas kinematics are essential to constrain the dark matter profile in galaxies. The cold gas kinematics excluded more than $95\%$ of models within the $1-\sigma$ confidence level of the Schwarzschild modelling with stellar kinematics only and reduced the relative uncertainty of the dark matter fraction to $10\%$ within $5 R_\mathrm{e}$. Adding the cold gas constraints does an excellent job on obtaining the inner slope $\gamma$ of the dark halo profile. 

\item We characterize the stellar orbits of NGC 2974 into three principal components: an extended hot component, a central warm component, and an extended cold component, corresponding to a prominent bulge, a central thick disc or a core, and a thin disc, respectively. As the cold gas kinematics are outside the field-of-view of the stellar kinematic data, the introduction of cold gas constraints does not alter the orbit distribution significantly.

\end{enumerate}

\section*{Acknowledgements}
The authors thank the referee for their helpful comments which helped to improve this manuscript.

MY gratefully acknowledges the financial support from China Scholarship Council (CSC) and Scottish Universities Physics Alliance (SUPA) for this project. MY thanks the Max Planck Institute for Astronomy, Heidelberg, for their hospitality during her working visit.

LZ acknowledges support from Shanghai Astronomical Observatory, Chinese Academy of Sciences under grant NO.Y895201009.

GvdV acknowledges funding from the European Research Council (ERC) under the European Union's Horizon 2020 research and innovation programme under grant agreement No 724857 (Consolidator Grant ArcheoDyn).

The Pan-STARRS1 Surveys (PS1) and the PS1 public science archive have been made possible through contributions by the Institute for Astronomy, the University of Hawaii, the Pan-STARRS Project Office, the Max-Planck Society and its participating institutes, the Max Planck Institute for Astronomy, Heidelberg and the Max Planck Institute for Extraterrestrial Physics, Garching, The Johns Hopkins University, Durham University, the University of Edinburgh, the Queen's University Belfast, the Harvard-Smithsonian Center for Astrophysics, the Las Cumbres Observatory Global Telescope Network Incorporated, the National Central University of Taiwan, the Space Telescope Science Institute, the National Aeronautics and Space Administration under Grant No. NNX08AR22G issued through the Planetary Science Division of the NASA Science Mission Directorate, the National Science Foundation Grant No. AST-1238877, the University of Maryland, Eotvos Lorand University (ELTE), the Los Alamos National Laboratory, and the Gordon and Betty Moore Foundation.



\bibliographystyle{mnras}
\bibliography{main}



\bsp	
\label{lastpage}
\end{document}